\documentstyle[psfig]{mn}  
\newif\ifAMStwofonts  
  
\newcommand{\apj}{\rm ApJ}

\newcommand{\apjs}{\rm ApJS}
\newcommand{\aaps}{\rm A$\&$AS}
\newcommand{\aap}{\rm A$\&$A}
\newcommand{\mnras}{\rm MNRAS}
\newcommand{\aj}{\rm Astron. J.}

\newcommand{\hb}{H$\beta$}

\newcommand{\etal}{~et~al.~}                                

\newcommand{\der}[2] {\frac{{\rm d}#1}{{\rm d}#2} }
\newcommand{\Msol}{M$_{\odot}$}                           

\ifoldfss  
  \ifCUPmtlplainloaded \else  
    \NewTextAlphabet{textbfit} {cmbxti10} {}  
    \NewTextAlphabet{textbfss} {cmssbx10} {}  
    \NewMathAlphabet{mathbfit} {cmbxti10} {} 
    \NewMathAlphabet{mathbfss} {cmssbx10} {} 
  \fi  
  \ifAMStwofonts  
    \ifCUPmtlplainloaded \else  
      \NewSymbolFont{upmath} {eurm10}  
      \NewSymbolFont{AMSa} {msam10}  
      \NewMathSymbol{\upi}     {0}{upmath}{19}  
      \NewMathSymbol{\umu}     {0}{upmath}{16}  
      \NewMathSymbol{\upartial}{0}{upmath}{40}  
      \NewMathSymbol{\leqslant}{3}{AMSa}{36}  
      \NewMathSymbol{\geqslant}{3}{AMSa}{3E}

    \fi  
  \fi  
\fi 
  
\ifnfssone  
  \newmathalphabet{\mathit}  
  \addtoversion{normal}{\mathit}{cmr}{m}{it}  
  \addtoversion{bold}{\mathit}{cmr}{bx}{it}  
  \newmathalphabet{\mathbfit} 
  \addtoversion{normal}{\mathbfit}{cmr}{bx}{it}  
  \addtoversion{bold}{\mathbfit}{cmr}{bx}{it}  
  \newmathalphabet{\mathbfss} 
  \addtoversion{normal}{\mathbfss}{cmss}{bx}{n}  
  \addtoversion{bold}{\mathbfss}{cmss}{bx}{n}  
  \ifAMStwofonts  
    \ifCUPmtlplainloaded \else  
      \UseAMStwoboldmath  
      \makeatletter  
      \new@mathgroup\upmath@group  
      \define@mathgroup\mv@normal\upmath@group{eur}{m}{n}  
      \define@mathgroup\mv@bold\upmath@group{eur}{b}{n}  
      \edef\UPM{\hexnumber\upmath@group}  
      \new@mathgroup\amsa@group  
      \define@mathgroup\mv@normal\amsa@group{msa}{m}{n}  
      \define@mathgroup\mv@bold\amsa@group{msa}{m}{n}  
      \edef\AMSa{\hexnumber\amsa@group}  
      \makeatother  
      \mathchardef\upi="0\UPM19  
      \mathchardef\umu="0\UPM16  
      \mathchardef\upartial="0\UPM40  
      \mathchardef\leqslant="3\AMSa36  
      \mathchardef\geqslant="3\AMSa3E  
    \fi  
  \fi  
\fi 
  
\ifnfsstwo  
  \DeclareMathAlphabet{\mathbfit}{OT1}{cmr}{bx}{it}  
  \SetMathAlphabet\mathbfit{bold}{OT1}{cmr}{bx}{it}  
  \DeclareMathAlphabet{\mathbfss}{OT1}{cmss}{bx}{n}  
  \SetMathAlphabet\mathbfss{bold}{OT1}{cmss}{bx}{n}  
  \ifAMStwofonts  
    \ifCUPmtlplainloaded \else  
      \DeclareSymbolFont{UPM}{U}{eur}{m}{n}  
      \SetSymbolFont{UPM}{bold}{U}{eur}{b}{n}  
      \DeclareSymbolFont{AMSa}{U}{msa}{m}{n}  
      \DeclareMathSymbol{\upi}{0}{UPM}{"19}  
      \DeclareMathSymbol{\umu}{0}{UPM}{"16}  
      \DeclareMathSymbol{\upartial}{0}{UPM}{"40}  
      \DeclareMathSymbol{\leqslant}{3}{AMSa}{"36}  
      \DeclareMathSymbol{\geqslant}{3}{AMSa}{\"3E}  
    \fi  
  \fi  
\fi 
  
\ifCUPmtlplainloaded \else  
  \ifAMStwofonts \else 
    \def\upi{\pi}  
    \def\umu{\mu}  
    \def\upartial{\partial}  
  \fi  
\fi  
  
  \title[How old are the HII Galaxies?]
{How old are the HII Galaxies?}
  
\author[Roberto Terlevich et al.]  
{Roberto Terlevich$^1$\thanks{RT and ET, Visiting Fellows at the Institute of Astronomy, 
Cambridge, U.K.}, 
Sergiy Silich$^1$, Daniel Rosa--Gonz\'alez$^{2}$ and   
Elena Terlevich$^1$ \\
$^1$ INAOE, Luis Enrique Erro 1. Tonantzintla, Puebla 72840. M\'exico.\\ 
$^2$ Astrophysics Group, Blackett Laboratory, Imperial College, Prince Consort Road, London SW7 2BW.\\  
}  

\date{Accepted  .  
      Received ;  
      in original form \today\ ~~  VERSION: v012}  
  
\pagerange{\pageref{firstpage}--\pageref{lastpage}}  
\pubyear{2003}  
  
\begin{document}  
  
\maketitle  
  
\label{firstpage}  
  
\begin{abstract}  

Using a novel approach we have reanalized the question 
of whether the extreme star forming galaxies
known as HII galaxies are truly young or rejuvenated old systems.

We first present a method of inversion that applies to any monotonic function
of time describing the evolution of independent events. We show that,  
apart from a normalization constant, the ``true'' time dependence can be 
recovered from the inversion of its probability density function.

We applied the inversion method to the observed  equivalent width of \hb\ (EW(\hb )) distribution  
for objects in the Terlevich and collaborators 
Spectrophotometric Catalogue of HII galaxies and found that 
their global history of star formation behaves much closer to 
the expectations of a continuos star formation model than to an 
instantaneous one. On the other hand, when the inversion 
method is applied to samples within 
a restricted metallicity range we find that their history of star formation 
behaves much closer to what the instantaneous model predicts.

Our main conclusion is that, globally, the evolution of HII galaxies seems 
consistent with a succession of short starbursts separated by quiescent periods 
and that, while the emission lines trace the properties of the
present burst, the underlying stellar continuum traces the whole star 
formation history of the galaxy. Thus, 
observables like the EW(\hb ) that combine an 
emission line flux, i.e. a parameter pertaining to the present burst, 
with the continuum flux, i.e. a parameter that traces the whole history 
of star formation, should not be used alone to characterize the present burst.

\end{abstract}  
  
\section{Introduction}

HII galaxies are dwarf emission line galaxies undergoing a burst of star
formation.  They are characterized by strong and narrow emission lines
originated in a giant star forming region
which dominate their {\em observable} properties at optical wavelengths.
Most HII galaxies are in fact Blue Compact
Galaxies (BCG's), but due to the different selection criteria
only a small percentage of BCG's,  i.e. those with the largest emission line 
equivalent widths, are HII galaxies.
We will stick to the name \,``HII galaxies\,'' to refer to the star forming
systems selected from objective prism surveys and  having strong narrow
emission lines.

Various studies of the
spectroscopic properties of HII galaxies in optical wavelengths
revealed systems of very low heavy
element abundances and high rates of star formation.
Earlier morphological studies have suggested that a large
proportion of the sample of HII galaxies observed are
compact and isolated  (Melnick 1987).  
This, together with the spectroscopic properties indicating low to very low 
metallicity
plus a very young stellar content, led
workers in the field ever since their discovery, to
question whether these systems
are truly young galaxies or made up by a few bursts separated by
long quiescent periods in the lifetime of the galaxy.
Reviews of the general properties of HII
galaxies can be found in  Melnick (1987), Terlevich (1988), 
the {\em \,``Spectrophotometric Catalogue
of HII Galaxies\,''}  (hereafter SCHG, Terlevich \etal 1991), 
Stasi\'nska \&  Leitherer (1996),
Telles \& Terlevich (1993; 1995; 1997), Telles, Melnick \& Terlevich (1997), 
and more recently in the excellent review by 
Kunth \& \"{O}stlin (2000).

More than 20 years ago Dottori (1981) suggested the use of the EW(\hb )
to detect differences in age among HII regions. Dottori \& Bica (1981) 
applied the 
EW(\hb ) method to 31 LMC and SMC HII regions and found an
uneven distribution of ages with most HII regions clustering at an
EW(\hb ) of about 70\AA\ and only 5 HII regions  with
EW(\hb ) $>$120 \AA . This result led them to suggest
that a galaxy-wide burst of star formation did occur 6 to 7 Myr ago
in the Magellanic clouds.

\begin{figure*}
\setlength{\unitlength}{1cm}           
\begin{picture}(7,8)         
\put(-5.5,-3.0){\includegraphics{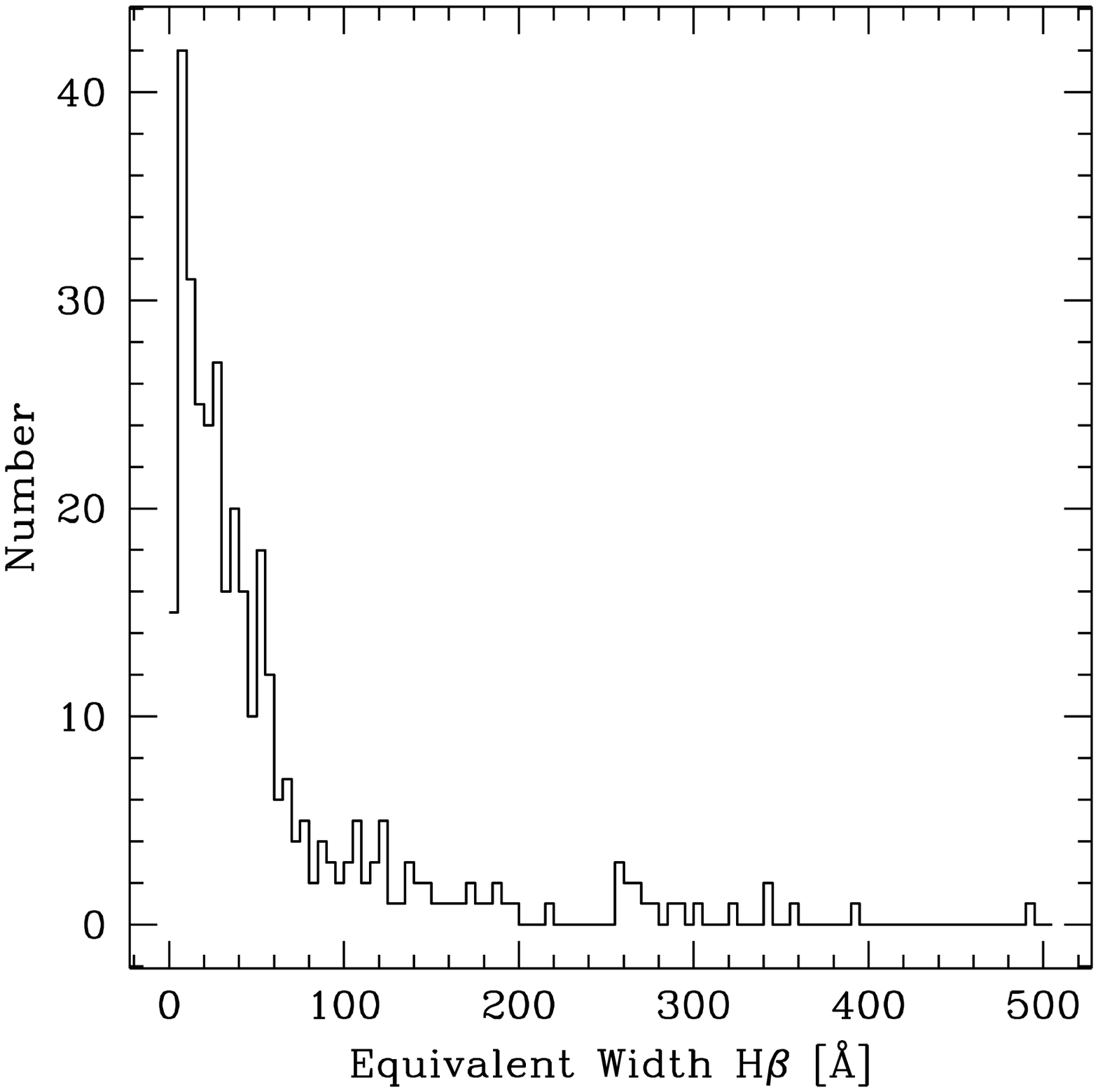}}
\put(3.5,-3.0){\includegraphics{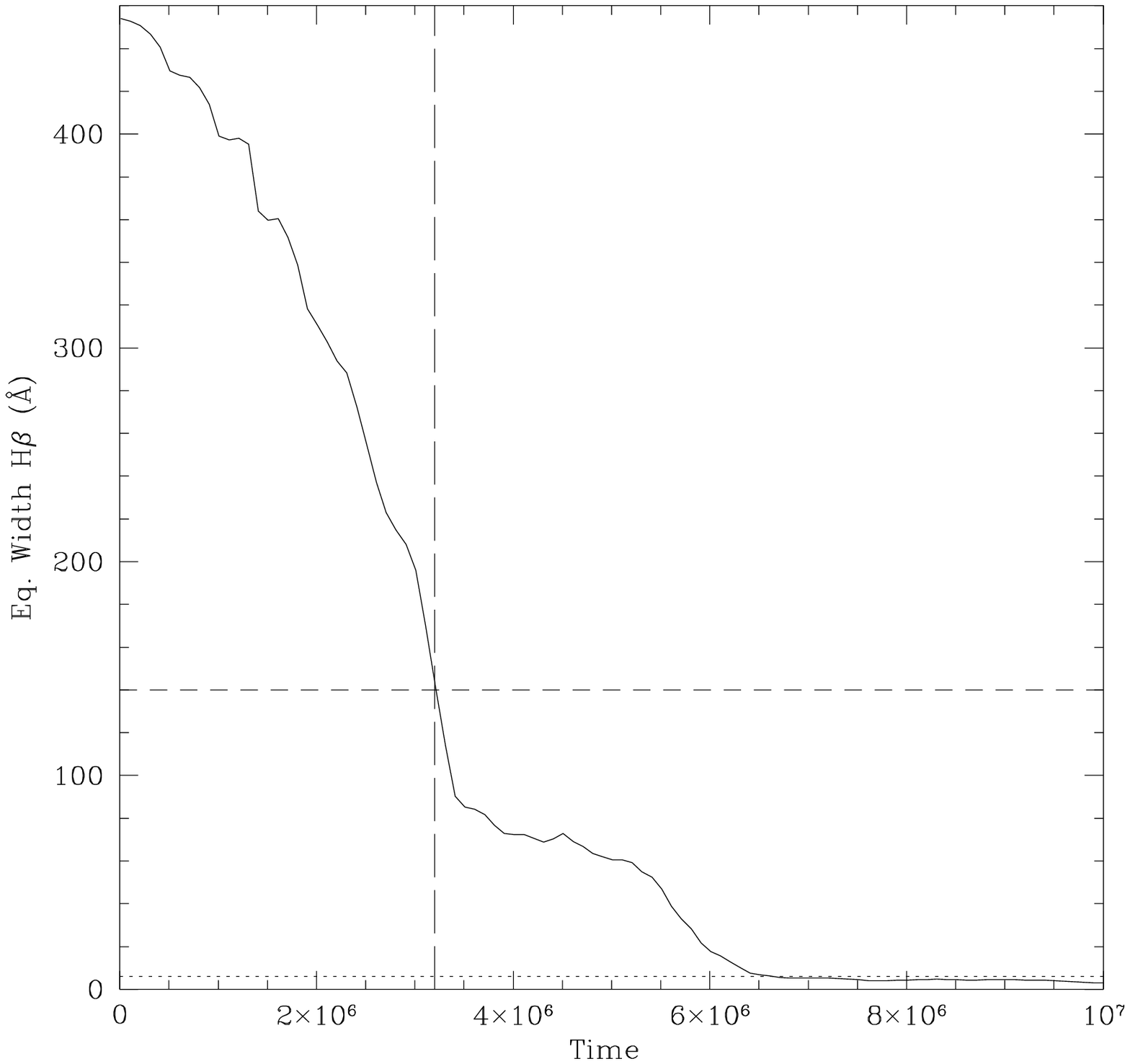}}
\end{picture}
\caption{
The equivalent width distribution of HII galaxies from 
the SCHG is plotted on the left. Clearly, and contrary to model 
expectations, there are only a few systems with EW(\hb ) $>$ 150\AA.
The predicted time evolution of a coeval population for a 
Salpeter IMF
with 100\Msol\ and 0.1\Msol\  upper and lower mass limit respectively is plotted on
the right. 
For a random population with a constant
production rate and considering that after 6.5 Myr a system will 
no longer be considered an HII galaxy,
about equal number of systems with ages below and above
3.2Myr are expected, i.e. about equal number of systems with 
EW(\hb ) above and below 150-200 \AA.
}
\label{figure1}
\end{figure*}

The SCHG compiles, among other parameters, line ratios
and equivalent widths for several hundred HII galaxies, and gives the
opportunity of investigating the ages of actively star forming galaxies to
find whether there is among them a truly young system. The SCHG is 
particularly
appropiate for this task because, due to the selection criterion used to find 
them, the HII 
galaxies in the SCHG are probably the youngest systems that can be studied 
in any detail.
This is due to the fact that the SCHG samples the narrow emission line 
galaxies with the highest
equivalent width in their emission lines, and is particularly biased
in the local universe
(at z$<$0.04) towards strong and compact emission in [OIII] 5007\AA.
This bias is introduced by the technique used in all the 3 sources of data,
the Cambridge, Tololo and University of Michigan surveys, that searched for 
strong emission using a  Schmidt camera (either the UK or the Tololo Schmidt
at the AAT and CTIO respectively) 
equipped with low dispersion objective prisms and IIIaJ emulsion, a combination
that produces a sensitivity range from 3500\AA\ to 5300\AA .

One of the early surprising results of the SCHG was that even in this strong emission 
line biased sample, there are less than 10$\%$ of the systems with EW(\hb ) $>$150 \AA\ (Fig.~1, 
left).
In contrast, as shown in Fig.~1, right, a quick analysis of the predictions for a single 
burst indicates that if star formation proceeeds uniformly with time, about half of the 
regions of star formation  should have an EW(\hb )
larger than about 150 - 200 \AA\ depending on model assumptions (all our 
synthesis computations use the SB99 code of Leitherer and collaborators 
(1999)).
  
Many explanations have been put forward for the lack of systems with high 
EW(\hb ). Chiefly among them, time dependent escape of ionizing photons,
dust affecting either the ionizing radiation or the visibility occurring
preferentially during
the first 3 Myrs, the presence of an underlying old population, 
uncertainties with the models, etc. The wide variety of possibilities
has had  the negative effect of almost halting the research in this 
important area.

We decided to take a fresh and different approach based as much as possible
on unsupervised analysis of samples.
We describe here a new method that, under the assumption
that the catalogue samples a population of bursting galaxies at different 
ages,
allows the reconstruction of the time evolution of the burst Balmer emission lines 
equivalent width, from their observed distribution.

\begin{figure*}
\setlength{\unitlength}{1cm}           
\begin{picture}(7,10.5)         
\put(-4.5,-0.3){\includegraphics{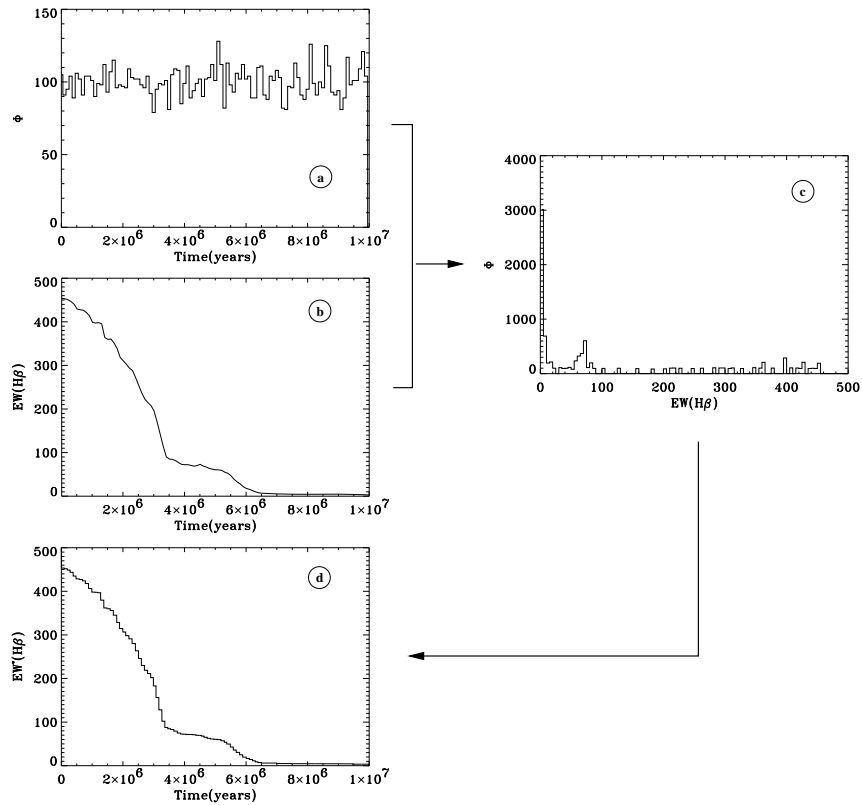}}
\end{picture}
\caption{Panel a shows the sample of random numbers 
homogeneously 
distributed in time, which we associate with the starburst birthrate. 
Panel b shows the adopted model, same as figure 1.
Panel c shows the computed distribution function of equivalent widths.
Panel d shows the inversion of panel c  distribution function using equation 
3. Note that the inversion closely reproduces the input model. 
}
\label{panels}
\end{figure*}


\section{The method}

\subsection{Brief description}

In this section we present our probability density distribution
inversion method. In contrast with other methods, we need only to
assume that the birth rate of starburts in the sample is random,
i.e. there is no relation in the ocurrence of starbursts in different galaxies,
and that the time evolution of the observed parameter is monotonic.

In fact, starbursts have a variety of ages, and hence a range of equivalent
widths of the emission lines. Two limiting cases are commonly used to describe 
their time evolution. 
They are the coeval starburst (SB) case, which assumes that all
stars are formed simultaneously in an instantanous SB
episode, where the characteristic time $\tau_{SB}$ is short compared to
the age of the galaxy ($\tau_{SB} \ll t$), and the  continuous star formation case, 
which assumes the star formation rate to be constant in time. The first 
case is widely applied for individual, moderate mass star clusters,
whereas the second one is assumed to be an average characteristic of the
massive star forming systems. 
In both cases the evolution of the EW(\hb ) is a monotonically declining 
function of time. The continuous star formation could also be
approximated as a sequence of small ``mini-bursts'' localized within
a rather small region in space and separated by short time intervals 
(see, e.g. Silich et al. 2002).

If the starbursts birth rate, R(t),
and the evolution of an individual starburst emission lines equivalent
width are known, then the probability distribution of starbursts 
with equivalent width is given by (e.g. Scalo \& Wheeler, 2001)
\begin{equation}
      \label{eq.5}
\rho_w(EW) = -\frac{1}{N_0} \frac{R[t(EW)]}{{\rm d}{EW}/{\rm d}t}, 
\end{equation}  
where N$_0$ is the total number of objects in the sample, and the 
function t(EW) describes the time evolution of the EW and is defined 
by the star formation mode.

Assuming a constant rate of star formation across the volume of the catalogue,
$$R(t) = N_0/t_{SF} = Const$$ 
one can obtain the normalized probability density as,
\begin{equation}
      \label{rhow}
\rho_w(EW) =  -\frac{1}{t_{SF}}{\der{t(EW)}{EW}}, 
\end{equation}  
where $t_{SF}$ is the total evolutionary time to be considered.

For the case of a monotonic time dependence
the inverse transformation of the EW distribution into a function EW(t)
is, from the relation (\ref{rhow}):
\begin{equation}
      \label{tew}
t(EW) = -\sum_{i_{max}}^{i} t_{SF} \, \rho_i(EW) \, \Delta EW =
     \frac{t_{SF}}{N_0} \sum_{i}^{i_{max}} \Delta N_i,  
\end{equation}  
where the limits i and i$_{max}$ correspond to the bins with equivalent
width EW and EW$_{max}$ respectively. 

Therefore, the shape of the time dependence of the
equivalent width can be recovered from the observed distribution.
On the other hand, the presence of an integration constant shows that
the characteristic star formation time scale $t_{SF}$ cannot be obtained
from the observed EW distribution alone.

\subsection{Numerical simulations}

We have run a number of tests on the analytical description above. Here we 
show the case for the evolution of the EW(\hb ) based on the expected
evolution of an instantaneous burst.

We first generated a sample of random numbers homogeneously 
distributed in time,  which we associate with the starburst
birthrate. For each value the corresponding equivalent width was 
calculated using the predicted evolution from SB99 
models. 
Panels a and b of figure \ref{panels} show the run of the burst rate and 
the adopted evolution of the EW(\hb ) respectively.
Panel c  shows the resulting probability density distribution 
function of equivalent widths. This represents the distribution of
EW(\hb ) in the homogeneously distributed instantaneous burst model catalogue.
Using equation 3 we inverted the catalogue EW probability density 
distribution function 
 shown in panel c and the reconstructed function is shown in panel d.
As expected from the analysis above, the inverted distribution function  
reproduces remarkably well the shape of the input time evolution of the 
equivalent widths.

\begin{figure*}
\setlength{\unitlength}{1cm}           
\begin{picture}(7,7.5)         
\put(-5.,-2.5){\includegraphics{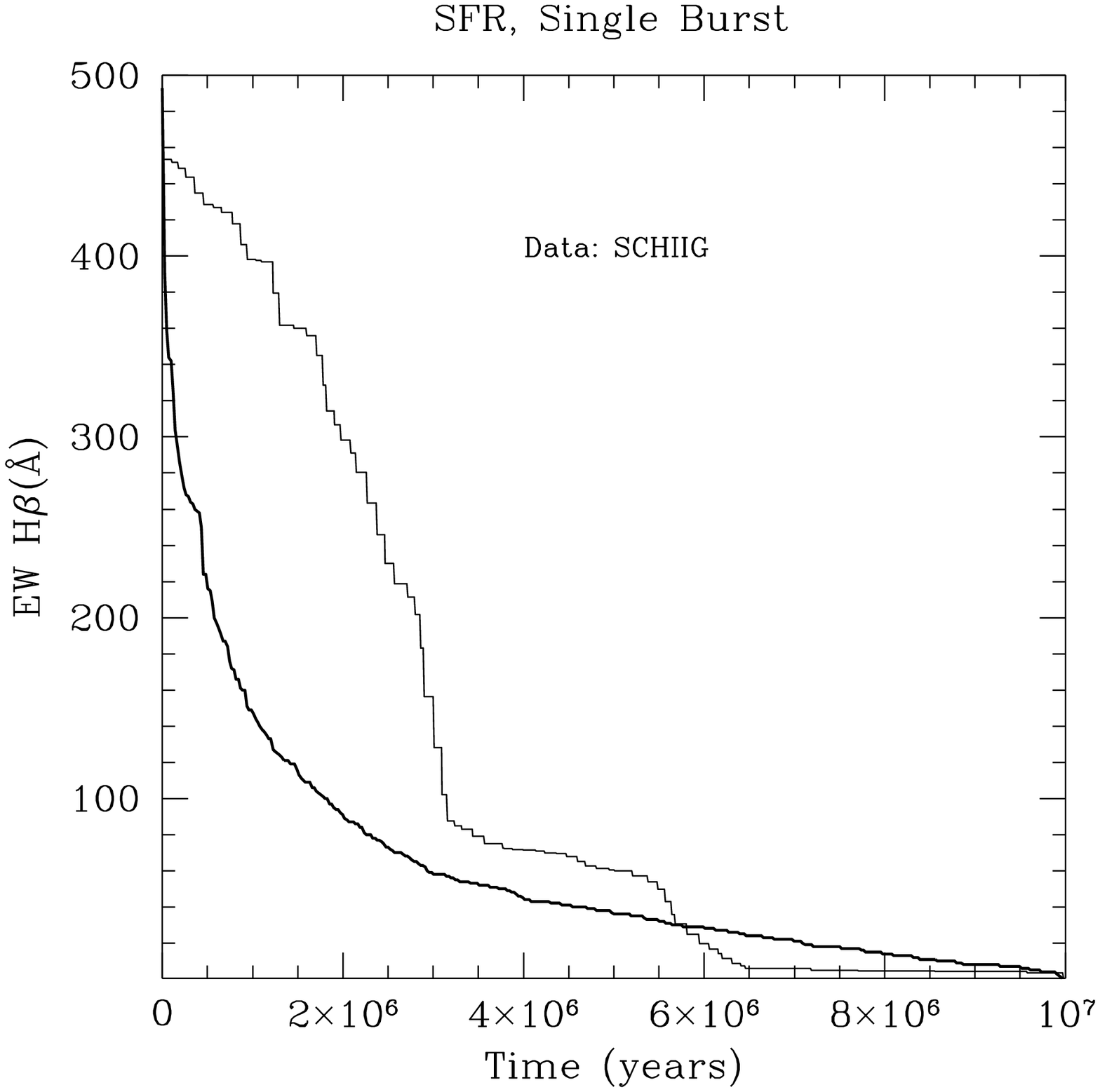}}
\put(4,-2.5){\includegraphics{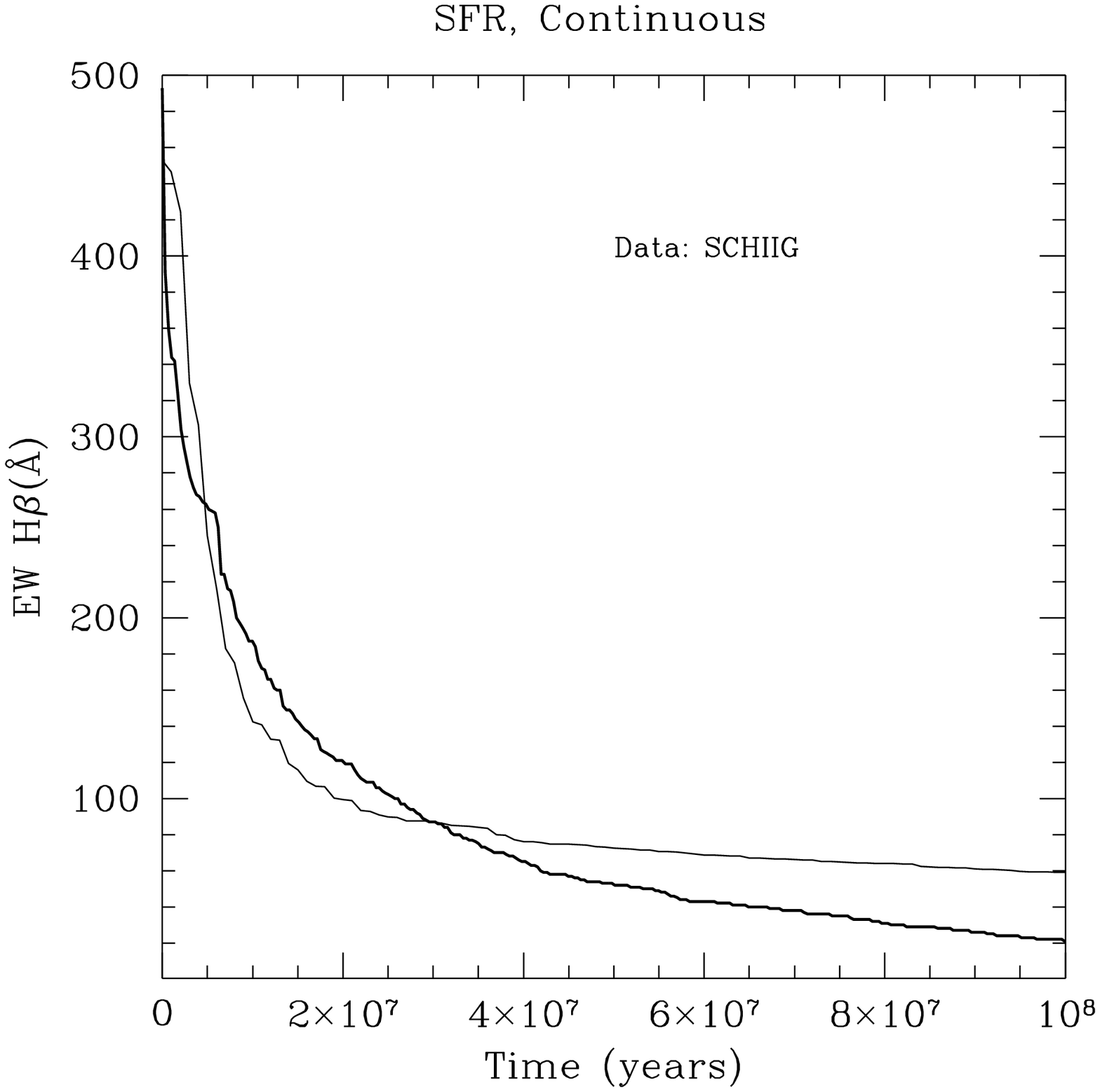}}
\end{picture}
\caption{Application of the inversion method to the SCHG. The thick line 
shows the inversion results
using equation 3 on the observed equivalent width distribution of HII 
galaxies (See
left panel of Fig.~1). The thin line shows the model predictions 
for an instantaneous burst (left panel) and for continuous star formation 
(right panel)}
\label{catalogue}
\end{figure*}

\section{Application to HII galaxies}

We have applied the inversion method to several samples of HII galaxies and 
giant HII regions.
We show here only the results for the SCHG that represents the more extreme 
case of bias towards strong line HII galaxies. The results for  other samples will be published 
elsewhere but the main conclusions of the present paper are unchanged 
by the extended sample analysis.


\begin{figure*}
\setlength{\unitlength}{1cm}           
\begin{picture}(7,11.5)         
\put(-3,-3.8){\includegraphics{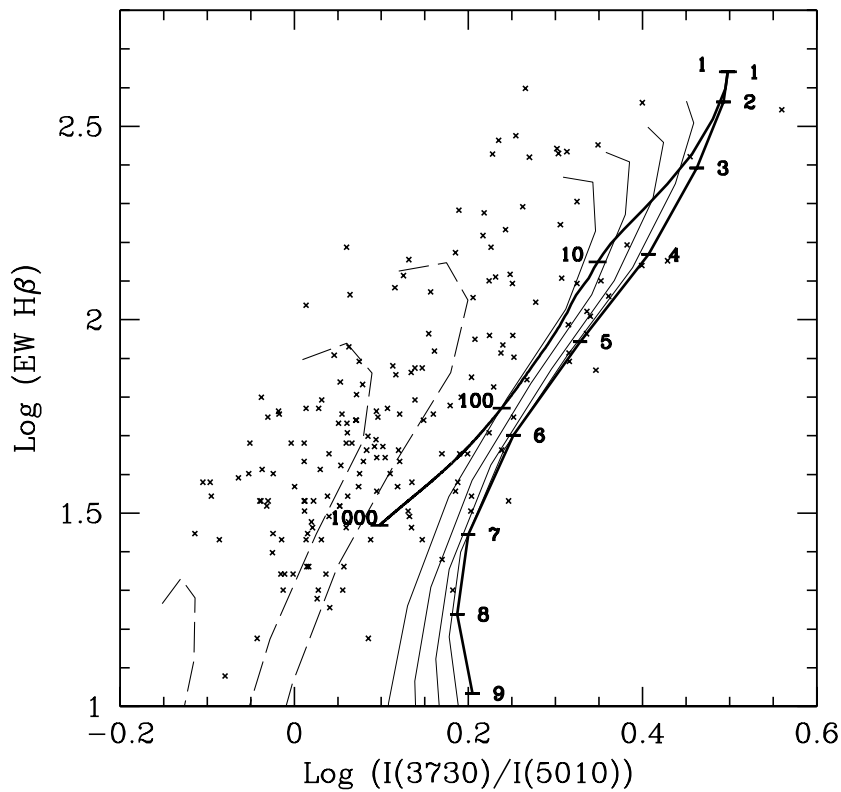}}
\end{picture}
\caption{The colour vs.~equivalent width plot for the models and the 217
HII galaxies from our sample.
The numbers on the model tracks represent the star cluster 
age (from 1Myr to 9Myr and from 1Myr to 1000Myr 
for the instantaneous and continuous star forming models, respectively).
The thin solid lines show the evolution of a system following a
sequence of identical bursts separated by 50 Myr time
intervals. For simplicity we plotted only 4 results, i.e. those starting at
50, 150, 350 and 750 Myr.
Dashed lines display evolutionary trends for the secondary 
starbursts occuring 900 Myr after the initial 
instantaneous  burst of star formation. The ratio of the secondary to the 
initial starburst mass drops from the right top to the left bottom lines 
like $10^{-2}$, 
$5 \times 10^{-3}$ and $10^{-3}$, respectively.
}
\label{colour2}
\end{figure*}

The results of the inversion are shown with the thick line in figure \ref{catalogue} while the thin 
lines show the model predictions 
for the instantaneous burst (left panel) and for the continuous star 
formation (right panel) cases.
It can be seen from figure \ref{catalogue} that the shape of the single burst 
model prediction is very different from that of 
the reconstructed evolution function of the EW(\hb).
The expected evolution of the EW(\hb) will be different for different initial conditions in the 
models like  the upper limit or slope of the IMF. 
We found that invariably all the predicted evolution curves have a convex shape (see figure
\ref{Zinversion} for different IMF values) while the inversion shows a concave
shape. In what follows we will use as reference for the model corresponding
to the instantaneous burst the prediction for $M_{up}=$120 \Msol\
and Salpeter slope. The reason  being that the sample is strongly 
biased towards high excitation
systems suggesting the need for  a combination of relatively
low metal content and a hot ionizing cluster to obtain it.  
This implies that stars more massive than 40-50 \Msol\ should be 
present at zero age.

On the right panel we can see that models with continuous star 
formation are in closer agreement with the shape of the 
time evolution of the EW(\hb ).
This rather surprising result perhaps indicates that the presence of older 
stars is affecting the continuum 
luminosity, consistent with earlier suggestions from, e.g.~Dufour et al.~(1996),
Garnett et al.~(1997), Legrand et al.~(2000).

\subsection{Continuum colour and age.}

To test on the somehow unexpected result of the previous section, 
we have analyzed other 
time dependent parameters that could provide independent information 
about the global age or evolutionary stage of HII galaxies.

The continuum colour is one of such age indicators. Particularly sensitive to 
early age are those colours that like Johnson's U-B, bridge the 
$\lambda\lambda$3800\AA\  to 4000\AA\ region.
For this analysis, we selected those HII galaxies
with the lowest dust reddenning correction (about 220 HII galaxies) and compared them
with SB99 estimates of the colours of an evolving stellar population.

In figure \ref{colour2} we have plotted the EW(\hb ) and  
the $\lambda\lambda$ 3730/5010\AA\ colour defined
as the ratio of the intensities of the adjacent continua to the 
[OII]$\lambda$3727\AA\ and [OIII[$\lambda$5007\AA\ 
emission lines.

Figure \ref{colour2} also shows in thick lines the evolution of the
instantaneous and  the continuous star formation models.
The digits along the curves represent the age of the models in units of Myr.

If HII galaxies were truly evolving as continuous star
forming systems, they will not depart much from the
continuous star formation line. Furthermore we will expect a symmetric distribution
with respect to the continuous star formation curve. 
This, however, is not the case for 
the sample of HII galaxies. Most of the observed values are above and/or to the
left of the continuous star formation model predictions.

In Figure \ref{colour2} we have also plotted the lines 
corresponding to multiple burst models. 
The thin solid lines represent the evolutionary 
sequence path for individual bursts
from a sequence of identical instantaneous starbursts, 
separated by 50 Myr quiescent 
intervals. The lines correspond to the bursts starting at 
50, 150, 350, and 750 Myr, respectively. 
The thin dashed lines show the evolution of a second burst
which occurs 900 Myr after the
initial one for a range of bursts mass ratios. 
The mass ratio of the
second to the first starburst changes from the right top
to the left bottom lines as 10$^{-2}$, $5 \times 10^{-3}$ and 
$10^{-3}$, respectively. 

It is possible to see that both the position and the scatter 
of the points in the EW(\hb ) vs.~colour diagram, 
are consistent with a population of galaxies undergoing multiple bursts 
of star formation during their cosmological evolution.
It is interesting to note the almost complete absence of points to the
right of the instantaneous case suggesting a good agreement
between models and observation. Reasuringly,  the highest observed 
EW(\hb ) is also consistent with the model predictions.

\subsection{Metallicity and age}

Another parameter that is expected to change in galaxies with 
age is their metal content. In particular, Oxygen that represents
more than 40 percent in mass of the metals, seems to be a good metallicity
indicator given that is mostly produced in massive stars with little
time delay (see, e.g. Pagel 1998).

\begin{figure*}
\setlength{\unitlength}{1cm}           
\begin{picture}(7,10.5)         
\put(-1.,-2.){\includegraphics{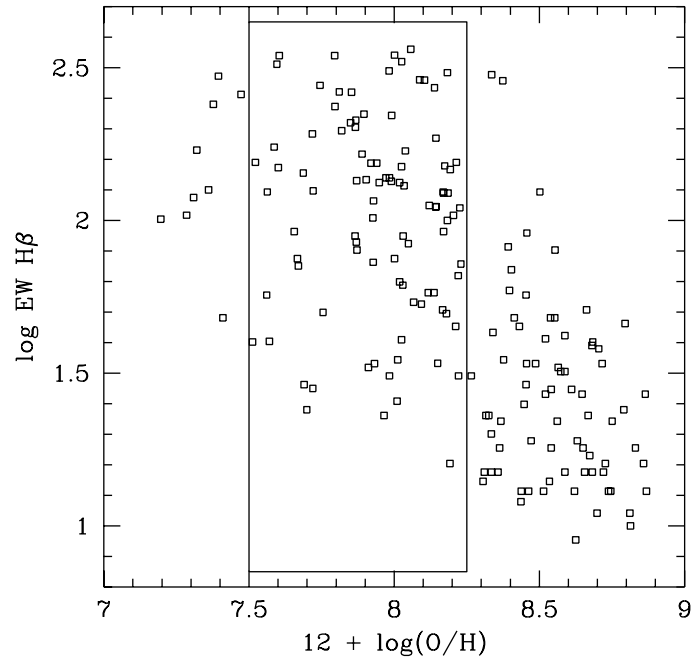}}
\end{picture}
\caption{Metallicity vs. EW(\hb ) from Denicol\'o et al.~(2002)
compilation. The rectangle isolates the selected restricted metallicity 
range used for the analysis (see text). It engulfes 70
objects.
}
\label{denicolo}
\end{figure*}

To investigate the behaviour of the metallicity of HII galaxies
versus their  EW(\hb) we have used the compilation of the best 
determinations of O/H in HII galaxies (Denicol\'o, Terlevich \& Terlevich 2002).
In figure \ref{denicolo} we have plotted metallicity and EW(\hb) for
the 183 star forming galaxies of  the Denicol\'o et al.~compilation.
Figure \ref{denicolo} shows a clear relation with EW(\hb ) 
albeit with a lot of scatter.
The tendency is in the sense that high metallicity HII galaxies 
show lower  EW(\hb ) while low metallicity HII galaxies have invariably
high EW(\hb ).

While a detailed analysis of the relation between metallicity and EW(\hb)
and its time evolution, requires galactic chemical evolution studies, 
outside the scope of this work and will be dealt with in a separate paper,
the inspection of simple multiple burst models (Pilyugin \& Edmunds 1996, Pilyugin 1999), 
nevertheless, allows as to predict that  by selecting a sample within a 
narrow range of metallicities, and  under the assumption that in short 
time scales, i.e. less than 10~Myrs, very little (if any) contamination 
due to the present burst occurs (e.g.~Roy and
Kunth, 1995), we may be able to construct a sample with a narrow
range in its chemical evolutionary age.
If this is the case, the inversion of the EW(\hb ) distribution of a sample
with a narrow metallicity range may recover a star formation history 
more closely related to a single event.

\begin{figure*}
\setlength{\unitlength}{1cm}           
\begin{picture}(7,9.7)         
\put(-1.,-2.){\includegraphics{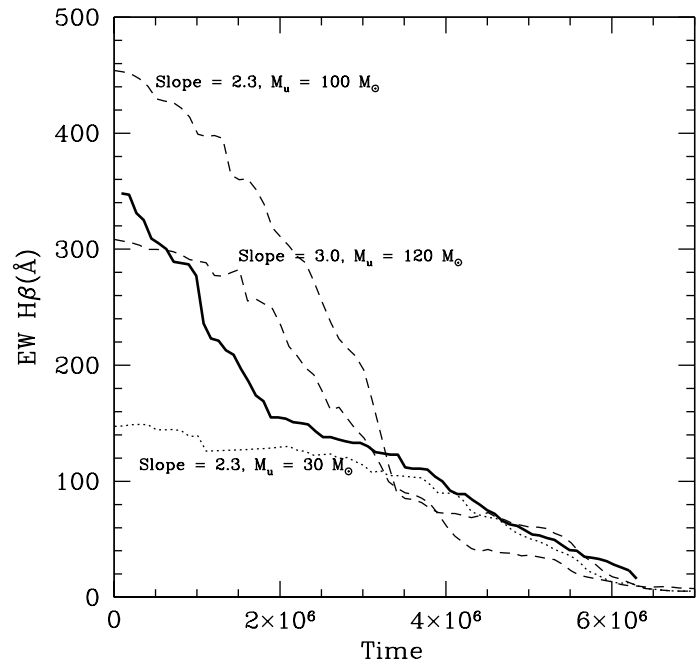}}
\end{picture}
\caption{The inversion of the time evolution of the EW(\hb ) for the 
restricted metallicity range subsample. The thick solid line represents the
inversion, and the thin dashed and dotted lines represent model predictions (SB99) with three
different upper mass values and two different slopes for the IMF, as labelled.
}
\label{Zinversion}
\end{figure*}

For this test, we have selected from Denicol\'o et al. (2002), the sub-sample of 70 galaxies 
covering the range
7.50 $<$ 12 $+$log (O/H) $<$ 8.25
The result of the inversion of the  distribution for this sample restricted in O/H 
is shown in figure \ref{Zinversion} together with the predictions from three
SB99 models. Two of the models represented in the 
figure, correspond to a Salpeter slope IMF having  upper mass 
limits of 100\Msol\ and 30\Msol\ 
respectively, the other corresponds to an upper mass of 120\Msol\ and 
an IMF slope of 3.0, i.e.~steeper than the Salpeter case. 

Clearly, the inversion has yielded an evolution whose shape is very close 
to that of the instantaneous models; furthermore the inversion seems 
more in agreement with an IMF with Salpeter slope and an upper  mass limit 
intermediate between 30\Msol and 100\Msol , or with a steeper slope IMF.

The simplest interpretation consistent with our findings is that
there are two different time scales for the evolution 
of HII galaxies on the metallicity - EW(\hb ) plane. 

On a time scale of about $10^7$yr
after any  starburst, the evolution on the metallicity - EW(\hb ) plane 
proceeds vertically downwards as it is
associated with a rapid decrease of the  EW(\hb ) as shown, e.g.~in 
figure (\ref{figure1}, right) and with basically no change in the metal 
content of the ionized gas. 

On time scales of order of $10^9$yr
there is the secular or cosmological evolution of the ISM metal
content which reflects on the stellar population build-up.

The superposition of these two time scales  results in the dispersion observed
in figures
\ref{colour2} and \ref{denicolo}.

Our findings are also consistent with the idea that the observed value of 
the EW(\hb) results from the emission produced
in the present burst superposed on the continuum generated by the present
burst PLUS all the previous star formation.

\section{Conclusions}

We have developed a simple inversion tool that allows to reconstruct
from observed probability density distributions of some monotonical
parameter, its time evolution.

We applied the inversion method to the EW(\hb ) distribution of a sample 
of 217 extreme star forming 
galaxies from the SCHG. We have  shown  that, considering the sample as a whole, its 
EW(\hb ) evolution is not well described by a coeval burst model, and that 
HII Galaxies seem to have a star formation
history that is closer to that predicted by a continuous star formation model.

The simplest interpretation is that while the observed emission
lines track the present burst, the underlying continuum contains
the whole history of star formation of the HII galaxy.

Even though HII Galaxies are selected by their strong emission lines, 
they are not truly young galaxies. Most of them have undergone substantial 
star formation probably during the previous 
100 - 1000 My to the present burst.

The situation changes when the analysis is
restricted to a sub-sample covering a narrow range in 
metallicities ($7.50<12+log\,O/H<8.25$). In this case
the EW(\hb ) evolution seems well described by a coeval burst model with
an IMF having an upper mass limit around 80~\Msol.

Clearly it would be very interesting to extend the method to the analysis of larger samples of 
galaxies like, e.g., that one provided by the Sloane Digital Sky survey.

\section{Acknowledgments}  

Sergiy Silich, Daniel Rosa Gonz\'alez and Elena Terlevich gratefully acknowledge financial support
from  
CONACYT, the Mexican Research Council, through research grants 
\#~36132-E and
\#~32186-E.
ET, RJT and DRG are grateful for the hospitality of the IoA, Cambridge, where 
part of this work was accomplished.

\bsp  
\label{lastpage}  
\end{document}